\documentclass[12pt]{article}
\usepackage{graphicx}
\usepackage{amsmath}

\DeclareRobustCommand{\greektext}{%
  \fontencoding{LGR}\selectfont\def\encodingdefault{LGR}}
\DeclareRobustCommand{\textgreek}[1]{\leavevmode{\greektext #1}}
\DeclareFontEncoding{LGR}{}{}
\DeclareTextSymbol{\~}{LGR}{126}
\title{\textbf{\large{Phase space parametrization of rain: the inadequacy of the gamma distribution}}}
\author{Massimiliano Ignaccolo\\  \textit{\footnotesize Dept. of Earth \& Ocean Sciences, Nicholas School of the Environment, Duke University,}
\\\textit{\footnotesize Durham, North Carolina} \and Carlo De Michele\\  \textit{\footnotesize DIIAR, Politecnico di Milano, Milano, Italy}} 

\begin{document}
\maketitle
\begin{abstract}
We show that the Gamma distribution is not an adequate fit for the probability density function of drop diameters using the Kolmogorov-Smirnov goodness of fit test. We propose a different parametrization of drop size distributions, which not depending by any particular functional form, is based on the adoption of standardized central moments. The first three standardized central moments are sufficient to characterize the distribution of drop diamters at the ground. These parameters together with the drop count form a 4-tuple which fully describe the variability of the drop size distributions. The Cartesian product of this 4-tuple of parameters is the rainfall phase space. Using disdrometer data from 10 different locations we identify invariant, not depending on location, properties of the rainfall phenomenon.
\end{abstract}

\section{Introduction }
At ``punctual'' space scale ($\sim$50 cm\textsuperscript{2}) rain 
can be described by a stochastic sequence of couples $(D_{j},\tau_{j})$:
$D_{j}$ being the diameter of the $j$-th drops, and $\tau_{j}$
the interval of time between the arrival of the $j$-th drop and the
$(j+1)$-th drop ($j=1,2,3,...$). Partitioning the time axis in sampling
time intervals of equal duration it is possible to build from the
sequence $(D_{j},\tau_{j})$ the sequence $(p_{k}(D),N_{k})$, with
$p_{k}(D)$ the probability density function of drop diameter, and
$N_{k}$ the number of drops observed at the ground during the $l$-th
sampling time interval ($l=1,2,....$). These two last quantities
are usually measured by disdrometers with a sampling time in the range
10s-5min. Observational evidences suggest that rain is a non-homogeneous
process as the variability observed in the sequence $(p_{l}(D),N_{l})$
cannot simply ascribed to the variability expected when sampling from
a single stochastic process \cite{Smith93,Lave_gole06,us-joh,us-hess1,us-hess2}.
The sequence $(D_{j},\tau_{j})$, and thus $(p_{l}(D),N_{l})$, is
non-stationary: the ``statistical rules'' to which the couples $(D_{j},\tau_{j})$
obey are not invariant under time translation \cite{Uhlenbeck_1930}.

A proper description of the distribution of drop sizes, which is fundamental
for understanding the microphysics involved in the mechanisms of precipitation
formation and retrieving rainfall from radar sensors, has been the
main research goal since the introduction of the impact disdrometer
has made reliable and prolonged measurements of drop diameters feasible.
The main tool the meteorological community uses to describe the variability
of the rainfall phenomenon is the drop size distribution, namely the
concentration $\mathcal{N}(D)$ per unit volume and diameter. In their
seminal work Marshall and Palmer \cite{MarshallPalmer48} proposed the exponential as
functional dependence for $\mathcal{N}(D)$ with a decaying constant
depending on the rainfall rate. Later, Joss and Gori \cite{JossGori78} made clear
that the exponential form is the results of sampling long time intervals
($\sim$30 minutes), and at smaller time intervals ($\sim$1 minute)
the drop size distribution does not have an exponential functional
dependence. Ulbrich \cite{Ulbrich83} proposed the Gamma distribution as
the functional form for the drop size distributions sampled at short
time intervals. The rationale for this choice is that while an exponential
decay is observed for the right tail of the distribution (recently
\cite{VillerBossa09} have shown that break-up due to air viscosity
is the main mechanism underlying the occurrence of exponential right
tail), the density at smaller diameters is smaller than that of an
exponential distribution. The Gamma distribution has been the only
choice adopted in literature (if one excludes sporadic attempts with
the log-Normal distribution, see e.g. \cite{Feingold:1986fk}, \cite{RosenfeldUlbrich03},
and \cite{Owolawi11}) mainly because of two properties: 1) it is
defined by two parameters, and 2) its moments can easily calculated
so that any rain bulk variable can easily expressed in terms of the
parameters of the Gamma distribution. The Gamma distribution has been
widely accepted by radar meteorology and cloud physics communities,
even if measurements show that the Gamma distribution is not general
enough to represent adequately the full range of the sample variability,
\cite{Klicheetal08}. The fit accuracy of the Gamma distribution
to disdrometer data has not been properly addressed (aside from subjective
statements: ``a good fit'' or ``a reasonable fit'' not supported
by any objective measurement) or mistakenly addressed (see Section
\ref{sec:Results}) as in \cite{smithjaetal09}. In the present work,
we consider data from 10 different sites, and use the Kolmogorov-Smirnov
goodness of fit test (e.g. \cite{Keutelian91,KottRoss97}) to show
that the Gamma distribution is a poor fit to 1 minute sampled drop
size distribution.

To obviate to this inadequacy, we propose a parametrization of the
drop size distribution based on a standard procedure in statistical
science: the description of an unknown probability density function
in terms of its mean $\mu$, standard deviation $\sigma$, skewness
$\gamma$, and kurtosis $\kappa$. In particular, we show that mean,
standard deviation and skewness are the minimum number of parameters,
which effectively describe the drop size distributions. These parameters,
together with the drop count $N$, are the variables necessary to
describe the rainfall phenomenon at a punctual scale in space and
at short time scales. All bulk variables of interest can be derived
by the 4-tuple of parameters $(N,\mu,\sigma,\gamma)$. Adopting the
jargon of the scientific community studying dynamical systems, we
refer to the four dimensional Cartesian product $\mathbf{R}^{4}$
spanned by the 4-tuple $(N,\mu,\sigma,\gamma)$ as the rainfall phase
space.

The work is organized as follows. In Section \ref{sec:Data}, we discuss
the data used in our analysis, Section \ref{sec:Methods} we illustrate
all the methods of analysis adopted. Section \ref{sec:Results} reports
our results, and Section \ref{sec:Conclusions} our conclusions.

\section{Data\label{sec:Data}}
We consider Joss-Waldvogel disdrometer data sampled at 1 minute time
intervals from ten different locations on Earth's surface. Table 1
gives the list of the locations with a three letter code used to reference
the site in the Figures. In addition, the Köppen-Geiger climate classification
(\cite{kotteletal06}) of each site is provided. Table 2 completes
the description of the data sets providing for each site the latitude,
longitude, altitude, together with number of minutes and drops considered
in the analysis.

\subsection{Data processing\label{sub:Data-processing}}
All data are processed as follows. We consider for each data base
only minutes for which the drop count is $\geq$60 (drop arrival rate
$\sim$1 per second). The rationale for this choice is twofold: A)
It guarantees a minimum reasonable accuracy for the estimation of
the probability density function $p(D)$, and the other statistical
parameters (mean, standard deviation, ...). B) It excludes time intervals
of observation which are quiescent (sparse precipitation) whose contribution
to the total cumulated precipitated volume is negligible (see discussion
in \cite{us-joh}). After this first filtering, we eliminate from
the remaining minutes of observation any outlier drop count. For each
minute, we find the disdrometer class with the maximum probability
density, and we calculate the central continuous non-zero span of
the probability density: the set of contiguous disdrometer classes
with non zero counts and which includes the class with maximum density.
The counts in all disdrometer classes which do not belong to the central
continuous non-zero span are considered as outliers and are discarded:
e.g. the disdrometer count (3,11,18,31,30,35,80,52,41,39,44,21,5,0,1,0,0,0,0,0)
has an outlier in the 15-th class which is disregarded leading to
the count (3,11,18,31,30,35,80,52,41,39,44,21,\\5,0,0,0,0,0,0,0). The
removal of outliers drop counts improves the estimate of higher moments
of the probability density function $p(D)$. A detailed discussion
on outliers and their effects on estimated statistical parameters
can be found in \cite{us-hess1}.

\section{Methods\label{sec:Methods} }

We now describe the methods used to quantify the variability of the
rainfall phenomenon.

\subsection{Variability of drop diameters}

Two equivalent descriptions are possible for the variability of drop
diameters. 1) The drop size distribution defined as the concentration
per unit volume and unit diameter $\mathcal{N}(D)$ 

\begin{equation}
\mathcal{N}(D)=N_{V}f(D),\label{eq:ND_proper}
\end{equation}
where $N_{V}$ is the number of drops per unit volume and $f(D)$
is the probability density function of drop diameter in the unit volume.
2) The flux-equivalent of Eq.(\ref{eq:ND_proper}): $N_{V}\rightarrow N$
and $f(D)\rightarrow p(D)$. $N$ is the number of drops observed
at the ground, and $p(D)$ is the probability density function of
drop diameter at the ground. The two descriptions are equivalent as

\begin{equation}
p(D)=\frac{{A_{m}Tv(D)}}{N}N_{V}f(D),\label{eq:ground-cloud}
\end{equation}
and 

\begin{equation}
N=A_{m}TN_{V}\intop_{0}^{\infty}v(D)f(D)dD.\label{eq:NandNV}
\end{equation}
In the above equations, $T$ is the time interval of observation (in
seconds), $A_{m}$ the capture area of the instrument (in m\textsuperscript{2}),
and $v(D)$ the drop velocity of diameter $D$ (in m/s). Usually it
is assumed that the arrival velocity of drops is equal to their limit
velocity and $v(D)=CD^{b}$, where $C=3.78$ m/(s mm\textsuperscript{0.67})
and $b=0.67$ (\cite{Ulbrich83}). With these definitions

\begin{equation}
p(D)=\Xi D^{0.67}N_{V}f(D)\;\;\;\;\;\;\Xi=\frac{{A_{m}TC}}{N}.\label{eq:cloud-ground-2}
\end{equation}
By use of Eq.(\ref{eq:cloud-ground-2}), we can connect the moments
$M_{\alpha,p}$ of the probability density function observed at the
ground with those of the concentration per unit volume $M_{\alpha,\mathcal{N}}$:

\begin{equation}
M_{\alpha,p}=\Xi M_{(\alpha+0.67),\mathcal{N}}\Longleftrightarrow M_{\alpha,\mathcal{N}}=\Xi^{-1}M_{(\alpha-0.67),p}.\label{eq:moments_connection}
\end{equation}
Using the above equation one is able to derive expressions for any
rain bulk variable. E.g. the rainfall rate $R$ (in mm/h) is

\begin{equation}
R=6\pi10^{-4}CM_{3.67,\mathcal{N}}=\frac{{6\pi10^{-4}}}{A_{m}T}NM_{3,p}.\label{eq:Rforboth}
\end{equation}
The concentration per unit volume, Eq.(\ref{eq:ND_proper}) is by
far the most common quantitiy in literature, even if measurements
of drop sizes at the ground are by large the only available data.
Hereby, we adopt the probability density of drop diameter at the ground,
Eq.(\ref{eq:ground-cloud}), as measured by disdrometer counts.

\subsection{The Gamma distribution for $\mathcal{N}(D)$ and $p(D)$}

The most common functional form adopted for the probability density
function $f(D)$ in Eq.(\ref{eq:ND_proper}) is the Gamma distribution 

\begin{equation}
f(D)=f_{\Gamma}(D,\lambda,k)=\frac{{\lambda^{k+1}}}{\Gamma(k+1)}D^{k}\exp(-\lambda D),\label{eq:defGamma}
\end{equation}
where $k$ is the shape,$\lambda$ the inverse scale parameter, and
$\Gamma(x)$ is Gamma function. Consequently the statistical moment
of order $\alpha$ of the drop size distribution $\mathcal{N}(D)$
is

\begin{equation}
M_{\alpha,\mathcal{N}}=N_{V}\int\limits _{0}^{\infty}dDD^{\alpha}f_{\Gamma}(D,\lambda,k)=N_{V}\frac{{\Gamma(k+\alpha+1)}}{\Gamma(k+1)}\lambda^{-\alpha}.\label{eq:momentsGamma}
\end{equation}
However, it is common practice, in Literature, to write Eqs.(\ref{eq:ND_proper})
and (\ref{eq:defGamma}) as

\begin{equation}
\mathcal{N}(D)=N_{0}D^{k}\exp(-\lambda D),\label{eq:ND_common}
\end{equation}
where $N_{0}$ is defined as 

\begin{equation}
N_{0}=N_{V}\frac{{\lambda^{(k+1)}}}{\Gamma(k+1)}.\label{eq:Nzero_NV}
\end{equation}
With the above notation

\begin{equation}
M_{\alpha,\mathcal{N}}=N_{0}\Gamma(k+\alpha+1)\lambda^{-(k+\alpha+1)}.\label{eq:momentsDSD_common}
\end{equation}
Finally, if the probability density function $f(D)$ is a Gamma distribution,
then also the probability density at the ground $p(D)$ is a Gamma
distribution

\begin{equation}
\mathcal{N}(D)=N_{v}f_{\Gamma}(D,\lambda,k)\Longleftrightarrow p(D)=f_{\Gamma}(D,\lambda,k+0.67).\label{eq:fromnuvolatoground}
\end{equation}

\subsection{Fitting the Gamma distribution to $\mathcal{N}(D)$ and $p(D)$}

We briefly review the two main methods which have been adopted in
literature to obtain the three parameters $(N_{V},\lambda,k)$ in
Eqs.(\ref{eq:ND_proper}) and (\ref{eq:defGamma}).

\subsubsection{Method of the Moments (MM)}

The method of moments (MM) uses the moments of the observed distribution
to derive the parameters of the desired fitting function. For a drop
size distribution with a Gamma distribution for diameter density,
the MM$_{\textrm{n1,n2,n3}}$ method finds the number of drops per
unit volume $N_{V}$, the scale $k$ and shape $\lambda$ such that
the resulting distribution $\mathcal{N}(D)$ exactly matches the moments
of order n1, n2 and n3 of the observed distribution. Hereby, we use
the MM$_{3,4,6}$ and MM$_{2,3,4}$ procedures
as they are the ones usually adopted in Literature (e.g. \cite{tokaishort96}
and \cite{testud01}), together with the MM$_{0.67,1.67,2.67}$ adopted
by \cite{smithjaetal09}. This last procedure finds the concentration
$N_{V}$, the scale $k$ and shape $\lambda$ of the drop size distribution
$\mathcal{N}(D)$ matching the observed number of drops at the ground
$N$ (moment 0.67), the observed average drop diameter at the ground
$\mu$ (moment 1.67), and the observed second moment of drop diameter
at the ground $M_{2,p}$ (moment 2.67). Note that the last two conditions
imply that the resulting drop size distribution matches the standard
deviation of drop diameter $\sigma$ observed at the ground. Therefore,
we will use for brevity the notation MM$_{N,\mu,\sigma}$
instead of MM$_{0.67,1.67,2.67}$.

Using Eqs. (\ref{eq:ND_proper}) and (\ref{eq:momentsGamma}), the
parameters $N_{V}$, $\lambda$, and $k$ are

\begin{multline}
\textrm{MM}_{3,4,6}\,\begin{cases}
N_{v} & =\frac{{\Gamma(k+1}}{\Gamma(k+5)}\lambda^{4}M_{4}\\
\lambda & =(k+4)\frac{{M_{3}}}{M_{4}}\\
k & =\frac{{11G-8+\sqrt{{G^{2}+8G}}}}{2-2G}\;\;\;\; G=\frac{{M_{4}^{3}}}{M_{3}^{2}M_{6}}
\end{cases}\label{eq:MM346}
\end{multline}

\begin{multline}
\textrm{MM}_{\textrm{2,3,4}}\begin{cases}
N_{v} & =\frac{{\Gamma(k+1)}}{\Gamma(k+2.67)}\lambda^{1.67}M_{1.67}\\
\lambda & =(k+3)\frac{{M_{2}}}{M_{3}}\\
k & =\frac{{-7G+\sqrt{{G^{2}+4G}}}}{2}\;\;\;\; G=\frac{{M_{3}^{2}}}{M_{2}M_{4}}
\end{cases}\label{eq:MM234}
\end{multline}

\begin{multline}
\textrm{MM}_{\textrm{N,\textgreek{m},\textgreek{sv}}}\,\begin{cases}
N_{v} & =\frac{{\Gamma(k+1)}}{\Gamma(k+2.67)}\lambda^{1.67}M_{1.67}\\
\lambda & =(k+1.67)\frac{{M_{0,67}}}{M_{1.67}}\\
k & =-4.34G+\sqrt{{G^{2}+4G}}\;\;\;\; G=\frac{{M_{1.67}^{2}}}{M_{0.67}M_{2.67}}
\end{cases}\label{eq:MM067167267}
\end{multline}

\subsubsection{Method of Maximum Likelihood (MML)}

The method of maximum likelihood is the main alternative to the method
of moments. Let $f_{X}\left(x,\mathbf{\theta}\right)$ be the probability
density function of the variable $X$ given the vector of parameters
$\mathbf{\theta}$, and $x_{1},\, x,...,\, x_{N}$ is a sample of
size $N$. The likelihood $L(\theta)$ that the sample $(x_{1},x_{2,},...,x_{n})$
is drawn from the distribution $f_{X}\left(x,\mathbf{\theta}\right)$
is defined as $L\left(\mathbf{\theta}\right)=\underset{i=1}{\overset{N}{\prod}}f_{X}\left(x_{i},\mathbf{\theta}\right)$.
In practice it is more convenient to deal with the logarithm of the
likelihood, denominated the log-likelihood $\ln L\left(\mathbf{\theta}\right)=\underset{i=1}{\overset{N}{\sum}\ln}f_{X}\left(x_{i},\mathbf{\theta}\right)$,
or the average log-likelihood $l\left(\theta\right)=\frac{1}{N}\ln L\left(\mathbf{\theta}\right)$.
The ML method makes an estimation of $\theta$ maximizing the average
log-likelihood, i.e. $\theta_{MML}=\underset{\theta}{arg\, max}\,[l\left(\theta\right)]$,
\cite{Edwards}. Dealing with disdrometer data, it is important to
recall that disdrometers collect drops with a diameter $D\geq D_{min}=0.3$
mm (for the JW disdrometer), making a lower truncation in the sample
distribution. In the absence of small drops in sample datasets, the
method of maximum likelihood ignoring this problem exhibits large
bias which do not decrease increasing the sample size, \cite{Klicheetal08}.
Consequently, modifications to the MML are necessary to deal explicitly
situations where lower truncations to the samples are present. We
will consider the lower truncated Gamma distribution which density
is
\begin{equation}\label{mml1}
f(D)=f_{\Gamma}(D,D_{min},\lambda,k)=\frac{\frac{{\lambda^{k+1}}}{\Gamma(k+1)}D^{k}\exp(-\lambda D)}
{1-\frac{\gamma\left(k+1,\lambda D_{min}\right)}{\Gamma\left(k+1\right)}},\: D>D_{min}
\end{equation}
where $\gamma\left(k+1,\lambda D_{min}\right)$ is the incomplete
Gamma function calculated in $\lambda D_{min}$. According to \cite{Johnsonetal09},
the average log-likelihood is
\begin{eqnarray}\label{mml2}
l\left(\lambda,k\right)=-\ln\left(1-\frac{\gamma\left(k+1,\lambda D_{min}\right)}{\Gamma\left(k+1\right)}\right)
+\left(k+1\right)\ln\lambda-\ln\Gamma\left(k+1\right)+\nonumber \\
+k\left(\frac{1}{N}\underset{i=1}{\overset{N}{\sum}\ln D_{i}}\right)
-\lambda\left(\frac{1}{N}\underset{i=1}{\overset{N}{\sum}D_{i}}\right).
\end{eqnarray}
The estimates of $\lambda$ and $k$ are obtained numerically minimizing
the function $l\left(\lambda,k\right)$, using the R code provided
by \cite{Johnsonetal09} in their appendix.

\subsection{Statistical characterization of a probability distribution function}\label{sub:Statpdf}

Under some general condition (e.g. \cite{reedsimon}), a probability
density function is completely determined by its moments: given the
sequence of moments $\{M_{j}\}$, $j=1,2,3,\dots$ there exists an
unique $f(x)$ such that $M_{j}=\int x^{j}f(x)dx$. This fact has
lead to the moment-characterization, in statistical sciences, of probability
density function for which a known parametric form is not available.
For this purpose a suitable number of moments and/or function of moments
will provide information on the unknown probability density function
$f(x)$ and, as a consequence, on the dynamical process driving the
realizations of the stochastic variable $x$. We refer to these parameters
as the \textit{statistical descriptors} of the probability density
function.

The two most commonly used statistical descriptors are the mean $\mu=M_{1}$,
and the standard deviation $\sigma=\sqrt{{M_{2}-(M_{1})^{2}}}$. In
addition to these two parameters, the skewness $\gamma$, measuring
the asymmetry of the distribution, and the kurtosis $\kappa$, measuring
the peakedness of the distribution, are used. Skewness and kurtosis
are the third and the fourth standardized central moments (the expectation
value of $[(x-\mu)/\sigma]^{3}$ and $[(x-\mu)/\sigma]^{4}$) and
can be written in terms of the moments of the distribution as follows:
\begin{equation}
\gamma=\frac{{M_{3}+2(M_{1})^{3}-3M_{1}M_{2}}}{[M_{2}-(M_{1})^{2}]^{3/2}}\label{eq:skewness-1}
\end{equation}
and
\begin{equation}
\kappa=\frac{{M_{4}-3(M_{1})^{4}+6(M_{1})^{2}M_{2}-4M_{1}M_{3}}}{[M_{2}-(M_{1})^{2}]^{2}}.\label{eq:kurtosis-1}
\end{equation}

Higher standardized moments do not have a particular name and are
generally not used since the higher the moment the larger are the
inaccuracies of any estimate. However, hereby we will make use of
the fifth standardized central moment, the expectation values of $[(x-\mu)/\sigma]^{5}$,
which will denote with the letter $\eta$
\begin{equation}
\eta=\frac{{M_{5}+4(M_{1})^{5}+10(M_{1})^{2}M_{3}-10(M_{1})^{3}M_{2}-5M_{4}M_{1}}}{[M_{2}-(M_{1})^{2}]^{5/2}}.\label{eq:5th-central-moment}
\end{equation}
Mean, standard deviation, skewness, and kurtosis are sufficient to
achieve a satisfactory (albeit not full) description of any probability
density function of interest. In many cases they are redundant as
some of the statistical descriptors are shown to be function of the
others. For example if $f(x)=f_{\Gamma}(x,\lambda,k)$ then only two
elements of the 4-tuple $(\mu,\sigma,\gamma,\kappa)$ are independent.
This redundancy also occurs for rainfall as we will show in Section
\ref{sec:Results}.

\subsubsection{Phase space\label{sub:Rainfall-phase-space}}

For a dynamical system, the term ``phase space'' indicates the Cartesian
product $\mathbf{R}^{n}$ of the $n$ variables necessary to describe
the system. Note that the cardinality $(n)$ of a dynamical system
is always larger or equal to its degrees of freedom. To describe the
motion a point particle of mass $m$ in one dimension we need its
position $x$ and its quantity of motion $p$ ($p=mv$, with $v=$velocity).
In this case the phase space is the $xp$ plane ($\mathbf{R}^{2}$),
and at each time $t$ the state of the particle is associated to a
point in the $xp$ plane. The time evolution of a dynamical system
is described by the time evolution (trajectory) of its associated
point in the phase space.

In the case of the rainfall phenomenon, the variables of interest
are bulk variables (e.g rainfall rate, reflectivity), which implies
a ``summation'' over the number of drops in a given interval of
time. All the bulk variables are functions of either the couple $(N_{V},f(D))$,
if one uses the concentration per unit volume, or the couple $(N,p(D))$,
if one uses the flux-equivalent description. If the probability density
function at the ground $p(D)$ (or that in a unitary volume $f(D)$)
has a parametric description: $p(D)=p(D,\theta_{1},\theta_{2},...,\theta_{m})$
in term of $m$ parameters, then we define the phase space of rainfall
as the Cartesian product $\mathbf{R}^{m+1}$ spanned by ($m$+1)-tuple
$(N,\theta_{1},\theta_{2},...,\theta_{m})$. E.g., if we consider
the concentration $\mathcal{N}(D)$ and assume a Gamma distribution
for the probability density function $f(D)$ in Eq.(\ref{eq:ND_proper}),
the rainfall phase space is $\mathbf{R}^{3}$ spanned by the 3-tuple
$(N_{V},\lambda,k)$. Hereby, we do not impose any particular functional
form on the function $p(D)$ (and thus $f(D)$). We show (Section
\ref{sec:Results}) that the parameters $\mu$, $\sigma$, and $\gamma$
are sufficient to describe the variability of the probability density
function at the ground $p(D).$ Therefore the Cartesian product $\mathbf{R}^{4}$
spanned by the 4-tuple $(N,\mu,\sigma,\gamma)$ can be considered
as the rainfall phase space.

Any bulk variable $B$ can be written as function of the phase space
parameters $B=B(N,\mu,\sigma,\gamma)$. In the case of the rainfall
rate $R$ (expressed in mm/h)

\begin{equation}
R=\frac{{6\pi10^{-4}}}{A_{m}T}N\bigg[\mu^{3}+3\mu\sigma^{2}+\sigma^{3}\gamma\bigg].\label{eq:R_mitica}
\end{equation}
 This equation states that the set of observation time intervals (1
minute in our case) with equal rainfall rate $R$ is a three dimensional
manifold in the phase space.

\subsection{Kolmogorov-Smirnov's goodness fit for probability distribution functions\label{sub:K-S test}}

The Kolmogorov-Smirnov (K-S) goodness-of-fit test is a non-parametric
test used to check if a sequence of random samples can be considered
as a realization of a stochastic process with a given cumulative distribution
function $F(x)$. The test compares the hypothetical $F\left(x\right)$
with the cumulative frequency $F_{N}\left(x\right)$, where $F_{N}\left(x\right)=i/\left(N+1\right)$
for $x_{\left(i\right)}\leq x<x_{\left(i+1\right)}$, $x_{\left(i\right)}$
is the i-th order statistics, and $i=1,..,N$. The K-S uses as test
statistic the maximum difference $D_{N}=max\left|F\left(x\right)-F_{N}\left(x\right)\right|$.
If no parameter in $F\left(x\right)$ is determined from data, then
$D_{N}$ has a distribution which is independent by $F\left(x\right)$.
Thus the critical value of $D_{N}$ for a significance level of $5\%$
and for large samples, $N>35$, is $1.3581/\sqrt{N}$, and reported
in all statistical textbooks (see e.g. \cite{KottRoss97}). Contrary,
if the parameters of $F\left(x\right)$ are estimated, then the distribution
of $D_{N}$ is dependent on $F\left(x\right)$, and the critical value
of $D_{N}$ must be re-calculated, e.g. via Montecarlo simulations
(\cite{Keutelian91}). The critical value of $D_{N}$ re-calculated
is always smaller than the value corresponding to the canonical case
where it is assumed that ``no parameter in $F\left(x\right)$ is
determined from data''. Disdrometer data report the occurrence of
a drop in a given range of diameter values (diameter class) and not
an ``exact'' diameter value which is needed to perform the Kolmogrov-Smirnov
test. To bypass this limitation we assign to a drop in the j-th diameter
class a random value selected uniformly in the range defined by the
class itself. 

\section{Results}\label{sec:Results}

All the results reported in this Section refer to probability density
function observed at the ground $p(D)$. Similar results can be obtained
for the concentration per unit volume and diameter $\mathcal{N}(D)$
since the two distribution are connected via Eq.(\ref{eq:ground-cloud}).

\subsection{Measuring the adequacy of the Gamma distribution}
The method of the moments MM$_{\textrm{n1,n2,n3}}$ for the Gamma
distribution finds the number of drops per unit volume $N_{V}$, the
scale $k$ and shape $\lambda$ such that the fitting concentration
per unit volume and unit diameter exactly matches the moments of order
n1, n2 and n3 of the observed concentration per unit volume and unit
diameter $\mathcal{N}(D)$. None of the MM$_{2,3,4}$, MM$_{3,4,6}$, and 
MM$_{N,\mu,\sigma}$ methods exactly match the moment 3.67, so that 
the ``reproducibility'' of the rainfall rate (observed versus the one derived from the fitted
parameters) has been considered as a ``measure'' of fit goodness:
e.g. (\cite{tokaishort96}) and \cite{smithjaetal09}). Hereby ,
we show that accuracy with which the MM$_{\textrm{n1,n2,n3}}$ method
reproduces the $j$-th moment of the concentration $\mathcal{N}(D)$
cannot be taken as a measure of fit goodness as the accuracy depends
on the separation between $j$ and the orders $n1$,$n2$,$n3$ and
not just on the particular functional form chosen. 

Let us consider the MM$_{3,4,6}$ method. The fitting distribution
by construction matches the third, fourth and sixth moment of $\mathcal{N}(D)$.
Therefore it is not surprising that the 3.67-th moment of the fitting
$\mathcal{N}(D)$ (the rainfall rate) is close to the observed value:
the middle-left panel of Figure 1 indicates the relative error $\chi$
is bounded in the range -0.5\%,0.5\%. However, instead of a Gamma
distribution for the functional form of $f(D)$, one could use any
other distribution with two parameters (e.g. Gaussian, Lognormal,
Beta) and obtain similar accuracies. What about the number of drops
observed at the ground $N$ (the 0.67-th moment of $\mathcal{N}(D))$?
As shown in the middle-right panel of Figure 1, the agreement is not
so good as relative error of the order of $\pm$25\% are possible.
Next, we consider the MM$_{N,\mu,\sigma}$
method. It reproduces the rainfall rate reasonably well, relative
error bounded in the -5\%,5\% range (upper-left panel of Figure 1),
but poorly reproduces the reflectivity $Z$ (the sixth moment of $\mathcal{N}(D)$)
as relative errors larger than $\pm$25\% are not so uncommon (upper-right
panel of Figure 1). Finally, the MM$_{2,3,4 }$ method reproduces
with the same accuracy of the MM$_{3.4.6}$ method (extremely
well) the rainfall rate, the relative error is bounded in the range
-0.5\%,0.5\% (bottom left panel of Figure 1), but is better than the
MM$_{3.4.6}$ method with respect the drop count $N$ (2
is closer to 0.67 than 3), relative error in the range -15\%,15\%
(bottom right panel of Figure 1). 

Figure 1 suggests that the Gamma distribution, as convenient and as
parsimonious it may be, is not a satisfactory functional form for
the drop size distribution. To prove this point we use a proper measure
of goodness-of-fit, such as the Kolmogorov-Smirnov test, and what
is considered to be the best fitting procedure, the MML method. In
Table 2, we report the percentage of acceptance (ACP) and rejection
(RJC) of the lower truncated Gamma distribution with parameters estimated
using the MML, using the Kolmogorov-Smirnov goodness-of-fit test with
a 5\% level of significance to each minute of the DRW data (6863).
In columns 2 and 3, the percentages are calculated using as 5\% critical
value $1.3581/\sqrt{N}$, the classical value reported in all statistical
textbooks (e.g. \cite{KottRoss97}) assuming that no parameters of
the Gamma distribution are estimated. In columns 4 and 5 the critical
value is determined via Montecarlo simulations taking into account
the fact that the parameters are estimated via the MML from data.
The first row (ALL) reports the fraction of the total number of minutes
in the DRW data sets for which the Gamma distribution can be considered
a good fit. The percentage of acceptance passes from 71\% to 45\%
when one takes in proper consideration that the parameters of the
distribution are obtained from the sample (\cite{Keutelian91}).
The remaining rows report the percentage of acceptance and rejection
for subsets obtained using as thresholds the percentiles of the distribution
of number of drops $N$ per minute: 5\%-, 25\%-, 50\%-, 75\%-, 95\%-percentile.
E.g. the second row reports the results for the subsets with number
of drops smaller than the 5\%-percentile ($N{}_{5\%}=66$ in our case):
minutes with a small sample size. On the other hand, the last row
reports the results for the subsets with number of drops larger than
or equal to the 95\%-percentile ($N{}_{95\%}=1589$ in our case):
minutes with a large sample size . We see that when we move to subsets
of minutes with large drop counts the percentage of acceptance (rejection)
diminishes (increases).

In summary, if we consider the entire DRW data set, we are confident
(at the 95\% level) that the Gamma distribution can be a proper fit
for probability density function $p(D)$ only for 45\% of 1 minute
sampling time intervals. More disturbingly the percentage of rejection
increases as the sample size increases. Note that test like the Kolmogorov-Smirnov
should be administered to sample with a size of at least $\sim$100
(\cite{KottRoss97}) to have of any significance (in other words
if the sample size is very small the effect of random fluctuations
is large enough that almost any tested distribution will pass the
test). Results for the other nine data sets (no reported here for
brevity) are similar to that of Darwin. On the ground of these results,
we reject the Gamma distribution as a proper fit for drop size distributions.

\subsection{Rainfall phase space}
We show that the value of standardized central moments of order $\geq$4
are strictly dependent from than the third one (skewness). In particular
we study the dependence on the skewness $\gamma$ of the fourth standardized
central moment (kurtosis $\kappa$) and the fifth one denoted by the
symbol $\eta$. As a consequence, mean ($\mu$), standard deviation
($\sigma$), and skewness ($\gamma$) provide a satisfactory description
of the variability of the probability density function $p(D)$ and
therefore the 4-tuple $(N,\mu,\sigma,\gamma)$ can be considered as
the rainfall phase space.

\subsubsection{higher standardized central moments}\label{sub:kappaetagamma}
To examine the dependence of the parameters $\kappa$ and $\eta$
on $\gamma$, we calculate for each data set the median, 5\%-, and
95\%-percentile of the observed values of $\kappa$ and $\eta$ for
a given value of $\gamma$ (in practice this is accomplished dividing
each data sets in subsets with ``equal'' ($\pm0.08$) value of skewness).
The results are reported in Figure 2. The median values are depicted
with solid lines of different colors one for each dataset. We also
calculate, merging all the datasets, the 5\% and 95\% percentile for
any given range value of the skewness if at least 100 samples (1 minute
time interval of observation) are present. The range between the 5\%
and 95\% percentile is shaded in gray in the figure. We see how in
both cases ($\eta$ vs $\gamma$, central panel and $\kappa$ vs $\gamma$
lower panel) the median lines are independent from the site chosen.
The discrepancies observed for values of skewness larger than $\sim$2.56
are mostly due to lack of statistics as shown in the upper panel of
Figure 2, where we plot the number of sample $M$ for each range value
of the skewness.

\subsubsection{Relationship between phase space parameters}\label{sub:musigmagamma}
Figure 2 shows that mean $\mu,$ standard deviation $\sigma$ and
skewness $\gamma$ are effective statistical descriptors for the probability
density function of drop diameters at the ground $p(D)$. Next, we
show that these three parameters are not independent. We divide the
range of value of the skewness in intervals of length 0.64 ({[}-1.60,-0.96{]},...,{[}2.88,3.52{]}
as in \cite{us-hess1,us-hess2}), and the values of the mean diameter
$\mu$ as follows: {[}0.3,0.4{]}, {[}0.4,0.5{]}, {[}0.5,0.6{]}, {[}0.7,0.8{]},
{[}0.9,1{]}, {[}1,1.2{]}, {[}1.2,1.4{]}, {[}1.4,1.6{]}, {[}1.6,1.8{]},
{[}1.8,2.0{]}, {[}2.0,2.5{]}, and {[}2.5,3{]} (indicated by the vertical
dashed lines in Figure 3). The rationale for these choices is to have
inside each range of skewness and mean a ``reasonable'' number (>10)
of 1 minute interval observations to calculate the median value of
the parameter $\sigma$. The results are reported in Figure 3. We
see how for value of the mean diameter less than 1 mm the median curves
are approximatively linear and do not depend on the site of observation.
In this case, the site average angular coefficient and average intercept
are reported at the bottom right of each panel, if at least the median
from 5 sites was available. We see how the slope (intercept) increases
(decreases) with increasingly larger value of the skewness until a
sort of plateau is reached for the skewness ranges $[1.60,2.24]$
and $[2.24,2.88]$ after which the slope decreases (intercept increases)
again. For values of the mean diameter larger than 1 the median curves
depend on the particular site of observation. However, the median
curve is also estimated from a smaller number of samples (in the range10-100).
Thus, with the present data, we cannot judge if the discrepancies
between sites for the median curve (in the range $\mu>1$) are real
properties of the rainfall phenomenon or merely artifacts due to the
poor statistical sampling available.

In summary, if $\mu$ is in the interval {[}0.3,1{]}
then
\begin{equation}
\sigma=a(\gamma)+b(\gamma)\mu\label{eq:musigmagamma}
\end{equation}
where $a$ is the intercept and $b$ the slope. Note that if the Gamma
distribution was indeed an extremely accurate fit to probability density
function of drop diameter at the ground p(D) then $\sigma=0.5\mu\gamma$
which is not supported by the experimental evidences depicted in Figure
3. Eq.(\ref{eq:musigmagamma}) suggests that only two parameters of
the triplets $(\mu,\sigma,\gamma)$ are necessary to describe the
probability density $p(D)$ of drop diameters at the ground. Thus
one could define the rainfall phase space as a tridimensional space:
e.g. the Cartesian product of the 3-tuple $(N,\mu,\gamma)$. However,
longer data sets are necessary to effectively estimate the functions
$a(\gamma)$ and $b(\gamma)$, and to explore the relationship $\sigma=a(\gamma)+b(\gamma)\mu$
in the range $\mu>1$mm. Therefore, for the purpose of this manuscript,
we conservatively consider $\mathbf{R}^{4}$ defined by the 4-tuple
$(N,\mu,\sigma,\gamma)$ as the rainfall phase space.

\subsubsection{Phase plots}
To each 1 minute time interval of observation is
associated the point of coordinates ($N,\mu,\sigma,\gamma$) in the
phase space. The entire data set occupies a volume in $\mathbf{R}^{4}$.
To visualize this volume, we need to consider the six 2D projections: 
$\mu-\log_{10}(N)$, $\sigma-\log_{10}(N)$, $\gamma-\log_{10}(N)$,
$\mu-\sigma$, $\mu-\gamma$, and $\sigma-\gamma$ (we use $\log_{10}(N)$
instead of $N$ for better visualization). Given a data set, we calculate
the density of points in the phase space for all six 2D projections.
Ten separate figures (one for each data set) would be necessary to
illustrate the results. To obviate to this difficulty and give an
idea to the reader of the differences/similarities between data sets
we adopted the concept of average bounding perimeter. For all 2D projections
we calculate the center of mass of the phase space points (each point
has the same mass). With the center of mass as fixed point we span
with an 10\textsuperscript{o} angle step the plane of the 2D projection.
For each 10\textsuperscript{o} cone we calculate the average distance
from the center of mass of the points within the cone. The connection
with a continuous line of all average distance creates the average
bounding perimeter which is a ``measurement'' of the volume of the
phase space occupied by the database. The results are shown in Figure
4. The plots on the $\mu-\sigma$ projection plane show how the average
bounding perimeters reflect the linear relationship between mean drop
diameter and standard deviation of drop diameter depicted in Figure
3. The plots on the $\mu-\log_{10}(N)$ plane projection indicate
that large value of counts ($\log_{10}(N)>2.8$) are reached (on average)
only for values of the mean drop diameter which are small ($\mu$
in the range 0.4-0.7 mm) or large ($\mu>1mm)$. The first case, many
drops with small diameter, is a common feature of orographic precipitation
as shown (e.g. \cite{us-hess1}, \cite{us-hess2}, \cite{martneretal08},
\cite{blanchard53}, and \cite{fujiwara67}). The BBY and CZC data
sets are the same ones used in (\cite{us-hess1}, \cite{us-hess2},
and \cite{martneretal08}) while BAO data set come from an instrument
located at 1,577 m asml. The second case, many drops with possibly
large diameters, is typical of strong convective events. the DRW data
sets (rain of monsoonic origin) and the MIK (small island at the equator)
and BKT (another equatorial site) are the data sets, among those considered,
where the combination of large number of drops and large drop diameters
occur more frequently on average. The average bounding perimeter on
the $\mu-\gamma$ plane projection show that a decrease in value of
the mean drop diameter is linked to a raise of the skewness value,
although this may be in part an effect of the limitation of the instrument
(Joss-Waldvogel impact disdrometer) which is not capable of detecting
drop diameters smaller than 0.3 mm (reducing the contribution of left
tails of $p(D)$ to the skewness). Due to the approximate linear relationship
between $\mu$ and $\sigma$, results of projection on planes for
which one of the axis is the standard deviation $\sigma$ are similar
to those for which axis is substituted by the mean $\mu$. If we consider
the result on the $\gamma-\log_{10}(N)$ plane projection, we see
how these two variable are quite uncorrelated as the shape of the
average bounding perimeters do not suggest any particular relation.
\subsubsection{Rain rate and phase space parameters}
The rainfall rate $R$ aside from a multiplicative constant is the
sum of three factors (Eq.\ref{eq:R_mitica}): $N\mu^{3}$, $3N\mu\sigma^{2}$,
and $N\sigma^{3}\gamma$. Each factor accounts for a fraction $\alpha\in[-1,1]$
(negative values are possible only for the factor $N\sigma^{3}\gamma$)
of the rainfall rate. We calculate $\alpha$ for each factor and each
1 minute time interval of observation. Then we calculate the probability
$F(\alpha)$ that the fraction does not exceed $\alpha$. The results
for each data base and each factor are shown in Figure 5. The $F(\alpha)$
curves are quite independent from the particular site. We see how
the factor $N\mu^{3}$ contributes the most to the rainfall rate with
a median contribution $\alpha_{m}$ ($F(\alpha_{m})=0.5$) which is
in the range 0.7-0.75. The second largest contribution comes from
the factor $3N\mu\sigma^{2}$, $\alpha_{m}$ in the range 0.2-0.25,
while the factor $N\sigma^{3}\gamma$ accounts for the smallest contribution:
$\alpha_{m}$in the range 0.025-0.05. These results can be explained
noticing that at all sites and for all time interval of observation:
1) $\sigma<1$, and 2) $\sigma<\mu$. So that $3\mu\sigma^{2}$ is
almost always smaller than $\mu^{3}$, and while $\gamma$ can be
larger (in absolute value) than $\mu$, $\sigma^{3}\gamma$ is always
smaller than $\mu^{3}$ in virtue of 1) and 2). 

\section{Conclusions}\label{sec:Conclusions}

When an objective measure of fit goodness is adopted, the Gamma distribution
provides a poor fit to the drop size distribution sampled at short
time scale (1 minute in our case) at all the ten sites considered.
It is the opinion of the Authors that only an objective criterion
of fit goodness (e.g. Kolmogorov-Smirnov) should guide the choice
of a particular functional form for the concentration $\mathcal{N}(D)$
and/or the probability density function at the ground $p(D)$. For
this reason we reject the Gamma distribution as a proper parametrization
of the rainfall phenomenon. We propose an alternative parametrization
based on the common statistical procedure of describing an unknown
probability density function in term of its standardized central moments.
We show that the 4-tuple of parameters $(N,\mu,\sigma,\gamma)$ is
sufficient to describe the observed variability of disdrometer counts
for all the ten sites considered, and refer to the Cartesian product
of 4-tuple $(N,\mu,\sigma,\gamma)$ as the rainfall phase space. The
volumes in the phase space relative to each data base (Figure 4) have
some common features and some discrepancies which reflect different
synoptic conditions and/or mechanisms of drop productions at play
at the different site considered. However some results remarkably
independent from the site considered: 1) standardized central moments
of order $\geq4$ have strong deterministic relationship with the
third standardized moment: the skewness. 2) mean $\mu$, standard
deviation $\sigma$, and skewness $\gamma$ are related to each other
via Eq.(\ref{eq:musigmagamma}) with values of the slope $b(\gamma)$
and intercept $a(\gamma)$ which are not compatible with a Gamma distribution
functional dependence. Finally, bulk variables of the rainfall can
be written as a function of 4-tuple $(N,\mu,\sigma,\gamma)$. e.g.
Eq.(\ref{eq:R_mitica}) for the rainfall rate $R$. Bulk variables,
such as the liquid water content $W$ and the reflectivity $Z$ are
proportional to fractional moments (2.33 and 5.33 respectively) of
the probability density function of drop diameter at the ground $p(D)$.
Therefore, analytical expressions for the variables $W$ ans $Z$
in terms of the 4-tuple $(N,\mu,\sigma.\gamma)$ are necessarily approximations,
on the contrary the Gamma distribution approximation leads to exact
analytical expression. However, the main result of this paper is that
any approximated expression, obtained via the proposed parametrization,
is physically meaningful while any exact expression, obtained via
the Gamma distribution parametrization, is not.

\section{Acknowledgment}
We wish to thank Dr.C.R.Williams and the National Oceanic and Atmospheric Administration (public availability of the 
data sets recorded at Bodega Bay, Eire, Cazadero, Darwin, and Kwajlein), the National Institute of Information and 
Communications Technology, Japan (Kashima data set), the Institute of Observational Research for Global Change together 
with the Japan Agency for Marine-Earth Science and Technology (Bukit Koto Tabang data set), Dr. Dan Brawn 
(Hassel data set), Dr. Martin Hagen of the Institut fuer Physik der Atmosphaere Deutsches Zentrum fuer Luft und 
Raumfahrt, Wessling, Germany (Macunaga data set), and the British Atmospheric Data Centre, Chilbolton data 
archive (Chilbolton data set).
\newpage
\begin{table}[t]
\caption{List of the sites from which Joss-Waldvogel disdrometer data are considered, with a three letters code for 
short referral and the K\"{o}ppen-Geiger climate classification.}
\begin{tabular}{|c|c|p{6cm}|}
\hline 
\textbf{Site} & \textbf{Code} & \textbf{K\"{o}ppen-Geiger climate classification}\tabularnewline
\hline 
\hline 
Eire, Colorado (USA) & BAO & Snow climate, fully humid with warm summer\tabularnewline
\hline 
Bodega Bay, California (USA) & BBY & Warm temperate climate with dry and warm summer\tabularnewline
\hline 
Bukit Koto Tabang, Indonesia & BKT & Equatorial rain forest, fully humid\tabularnewline
\hline 
Chilbolton, United Kingdom & CHB & Warm temperate climate, fully humid with cool summer and cold winter\tabularnewline
\hline 
Cazadero, California (USA) & CZC & Warm temperate climate with dry and warm summer\tabularnewline
\hline 
Darwin, Australia & DRW & Equatorial savannah with dry winter\tabularnewline
\hline 
Hassel, Germany & HSL & Warm temperate climate, fully humid with warm summer\tabularnewline
\hline 
Kashima, Japan & KSH & Warm temperate climate with dry and warm summer\tabularnewline
\hline 
Macunaga, Italy & MAC & Snow climate, fully humid with cool summer and cold winter\tabularnewline
\hline 
Kwajalein Atoll, RMI & MIK & Equatorial rainforest, fully humid\tabularnewline
\hline 
\end{tabular}
\end{table}

\begin{table}[t]
\caption{List of the sites (short code referral) from which Joss-Waldvogel disdrometer data considered with latitude, longitude, altitude, number of 1 minute time interval in data set, and total number of drops in the data set. The symbol ({*}) indicates quantities calculated after data sets are processed according to the procedure described in Section 2.1.}
\begin{tabular}{|c|c|c|c|c|c|}
\hline 
\textbf{Code} & \textbf{Long.} & \textbf{Lat.} & \textbf{Alt. (m)} & \textbf{\#min\textsuperscript{\textbf{{*}}}} & \textbf{\#drop\textsuperscript{\textbf{{*}}}}\tabularnewline
\hline 
\hline 
BAO & 40.05N & 105.00W & 1,577 & 6,016 & 2,349,280\tabularnewline
\hline 
BBY & 38.20N & 123.00W & 12 & 10,804 & 5,389,240\tabularnewline
\hline 
BKT & 0.12S & 100.19E & 864 & 68,389 & 25,109,376\tabularnewline
\hline 
CHB & 51.14N & 1.43W & 82 & 29,122 & 7,015,480\tabularnewline
\hline 
CZC & 38.61N & 123.22W & 475 & 76,137 & 44,252,384\tabularnewline
\hline 
DRW & 12.45S & 130.83E & 12 & 6,863 & 2,753,037\tabularnewline
\hline 
HSL & 51.51N & 7.1E & 60 & 26,402 & 7,072,649\tabularnewline
\hline 
KSH & 35.95S & 140.65E & 45 & 68,570 & 19,752,935\tabularnewline
\hline 
MAC & 45.97N & 7.96E & 1,300 & 9,956 & 3,275,264\tabularnewline
\hline 
MIK & 8.71N & 167.73W & 1 & 20,170 & 7,594,915\tabularnewline
\hline 
\end{tabular}
\end{table}

\begin{table}[t]
\caption{Percentage of acceptance and rejection of the (lower truncated) Gamma distribution using the Kolmogorov-Smirnov goodness-of-fit test to each minute of the DRW data (6863) with a 5\% level of significance. In columns 2 and 3 the critical value used is $1.3581/\sqrt{N}$ assuming that no parameters of the Gamma distribution are estimated, while in columns 4 and 5 the critical value is determined via Montecarlo simulations taking into account the fact that the parameters of Gamma are estimated via the MML from data. The results refer to entire DRW dataset (first row), and to subsets obtained considering as thresholds, the percentiles of the distribution of number of drops $N$ per minute (second tttlast row). The thresholds are $N{}_{5\%}=66$, $N{}_{25\%}=104$, $N{}_{50\%}=182$, $N{}_{75\%}=455$, and $N{}_{95\%}=1589$.}
\begin{tabular}{|c|c|c|c|c|}
\hline 
 & ACP & RJC & ACP & RJC\tabularnewline
\hline 
\hline 
ALL & 71\% & 29\% & \textbf{45\%} & \textbf{55}\%\tabularnewline
\hline 
$<N{}_{5\%}$ & 94\% & 6\% & \textbf{78\%} & \textbf{22\%}\tabularnewline
\hline 
$\geq N{}_{5\%}$ & 70\% & 30\% & \textbf{44\%} & \textbf{56\%}\tabularnewline
\hline 
$<N{}_{25\%}$ & 93\% & 7\% & \textbf{71\%} & \textbf{29\%}\tabularnewline
\hline 
$\geq N{}_{25\%}$ & 64\% & 36\% & \textbf{37\%} & \textbf{63\%}\tabularnewline
\hline 
$<N{}_{50\%}$ & 91\% & 9\% & \textbf{65\%} & \textbf{35\%}\tabularnewline
\hline 
$\geq N{}_{50\%}$ & 52\% & 48\% & \textbf{25\%} & \textbf{75\%}\tabularnewline
\hline 
$<N{}_{75\%}$ & 84\% & 16\% & \textbf{56\%} & \textbf{44\%}\tabularnewline
\hline 
$\geq N{}_{75\%}$ & 34\% & 66\% & \textbf{13\%} & \textbf{87\%}\tabularnewline
\hline 
$<N{}_{95\%}$ & 74\% & 26\% & \textbf{47\%} & \textbf{53\%}\tabularnewline
\hline 
$\geq N{}_{95\%}$ & 24\% & 76\% & \textbf{5\%} & \textbf{95\%}\tabularnewline
\hline 
\end{tabular}
\end{table}

\begin{figure}[t]
  \noindent\includegraphics[width=1\linewidth]{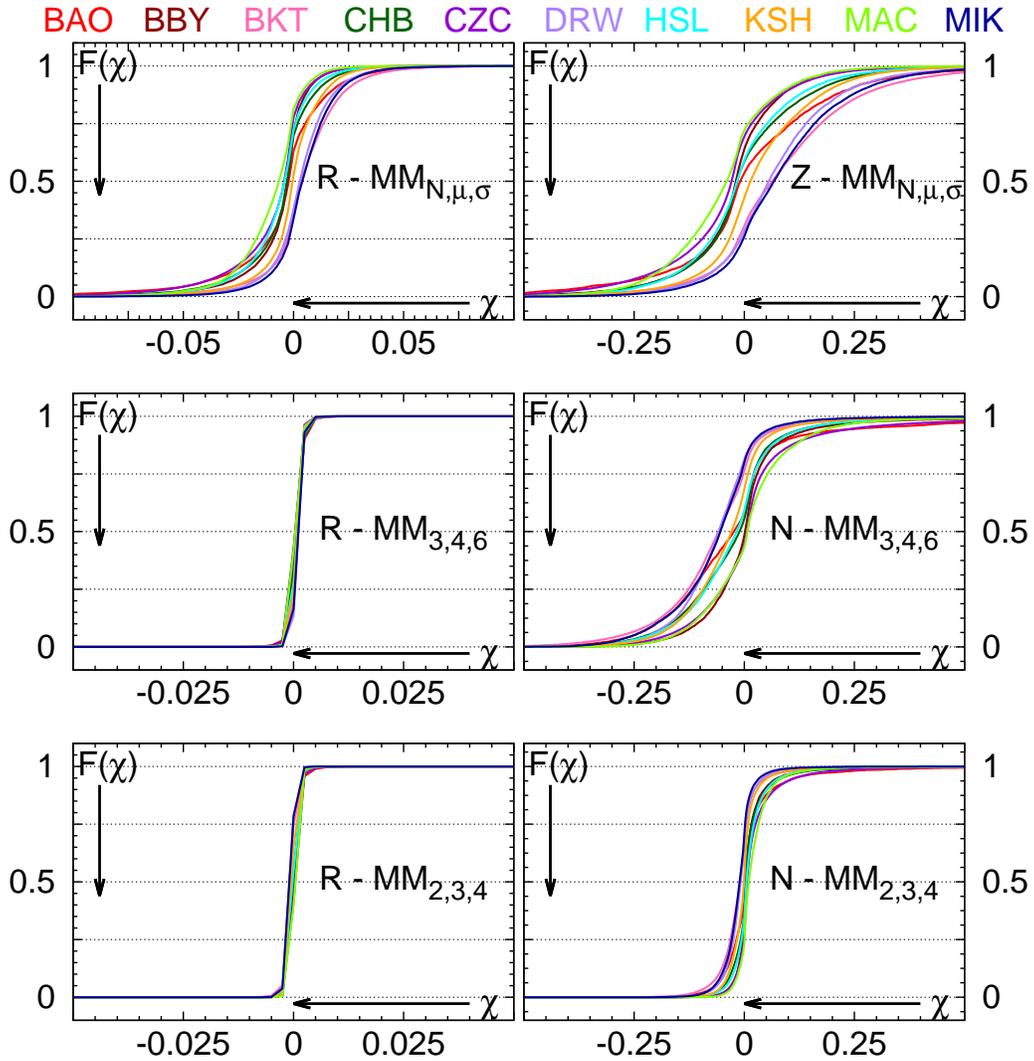}
\caption{The observed frequency $F(\chi)$ of exceeding the relative error $\chi$ when calculating different bulk 
variables with the fitted Gamma distribution to the probability density function $p(D)$ of drop diameters at the 
ground. Variables $R$ and $Z$ for the MM$_{N,\mu,\sigma}$ method (upper panels), 
variables $R$ and $N$ for the MM$_{3,4,6}$method (middle panels), and MM$_{2,3,4}$ method (bottom
panels). Different colors indicate different data bases.}\label{f1}
\end{figure}

\begin{figure}[t]
  \noindent\includegraphics[width=1\linewidth]{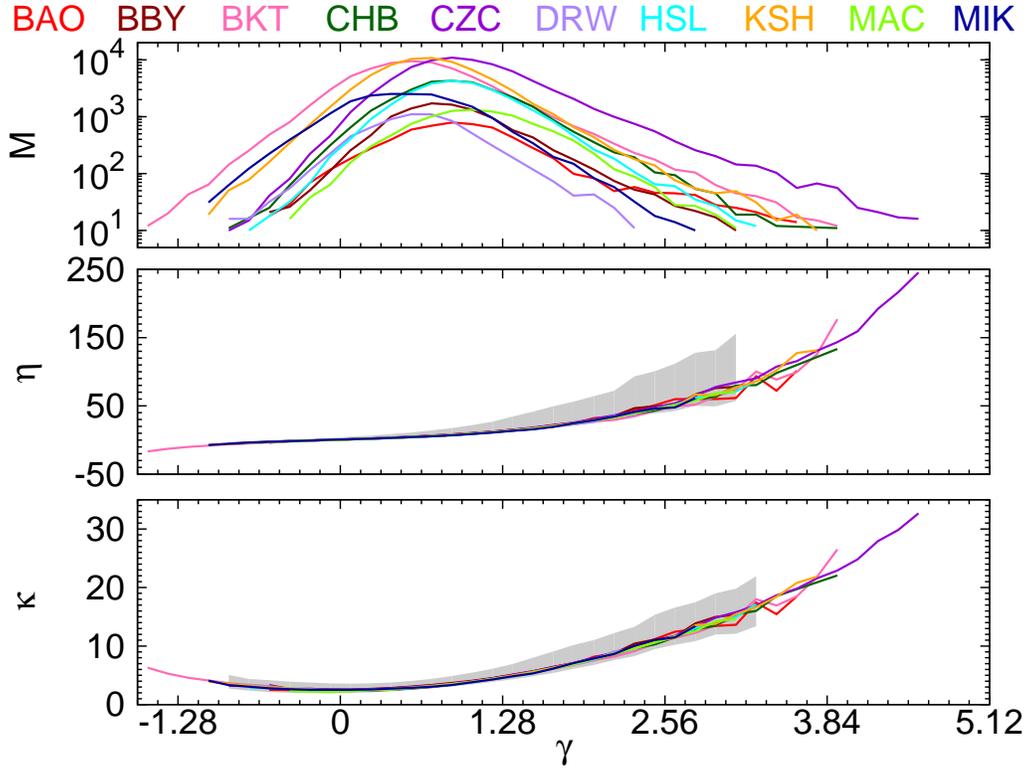}
  \caption{The number of samples $M$ (upper panel), the kurtosis $\kappa$ (middlepanel), and the fifth standardized 
central moment $\eta$ (lower panel)as a function of the skewness $\gamma$. Solid lines indicate the number of samples 
and the median values for the parameters $\kappa$ and $\eta$. Gray shadowed areas represent the 5\%-, 95\%-percentile
range. Different colors indicate different data bases.}\label{f2}
\end{figure}

\begin{figure}[t]
  \noindent\includegraphics[width=1\linewidth]{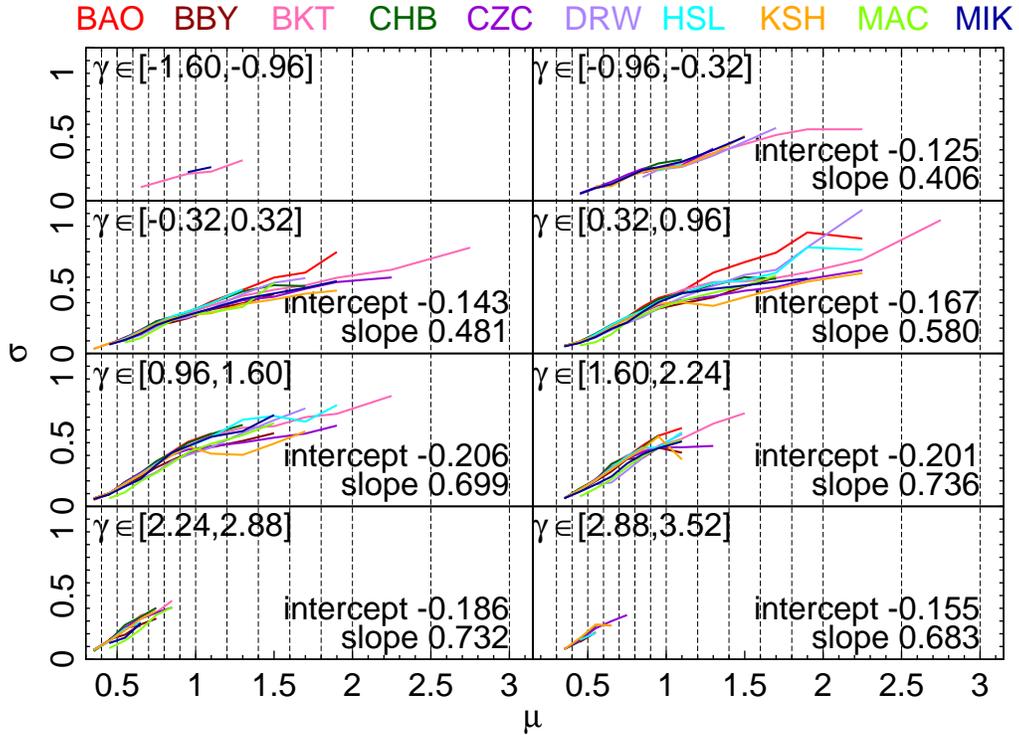}
  \caption{The median values of the standard deviation $\sigma$ of drop diameters for different range value of the mean 
diameter $\mu$ (indicated by vertical dashed lines) and different range values of the skewness of drop diameter $\gamma$ 
(shown in the top left corner of each panel). Also shown in the bottom right corner of each panel is the site average
slope and intercept of the median standard deviation curves in the interval $\mu\in[0.3:1]$. Different colors indicate 
different databases.}\label{f3}
\end{figure}

\begin{figure}[t]
  \noindent\includegraphics[width=1\linewidth]{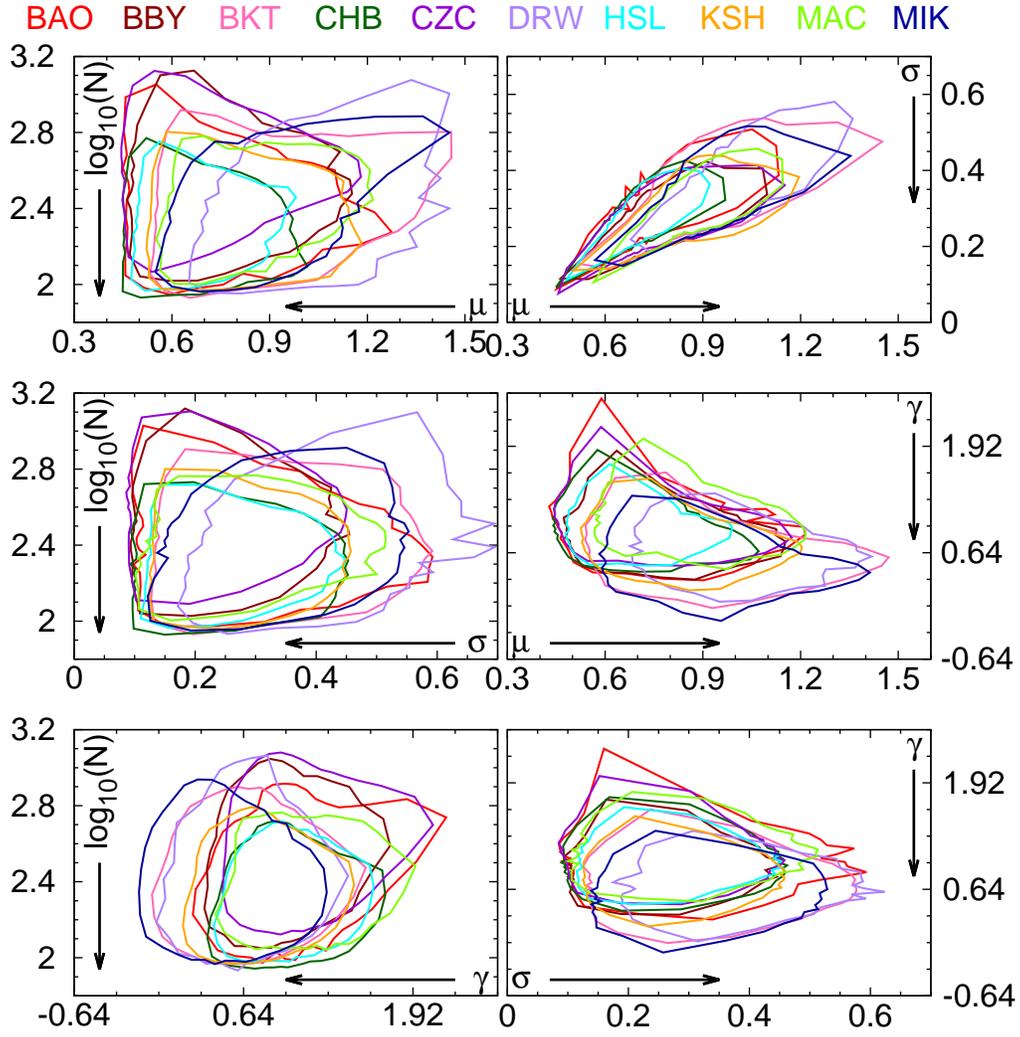}
  \caption{Phase space 2D projections plots of the average bounding perimeter lines for each site of observation. 
Different colors indicate different data bases.}\label{f4}
\end{figure}

\begin{figure}[t]
  \noindent\includegraphics[width=1\linewidth]{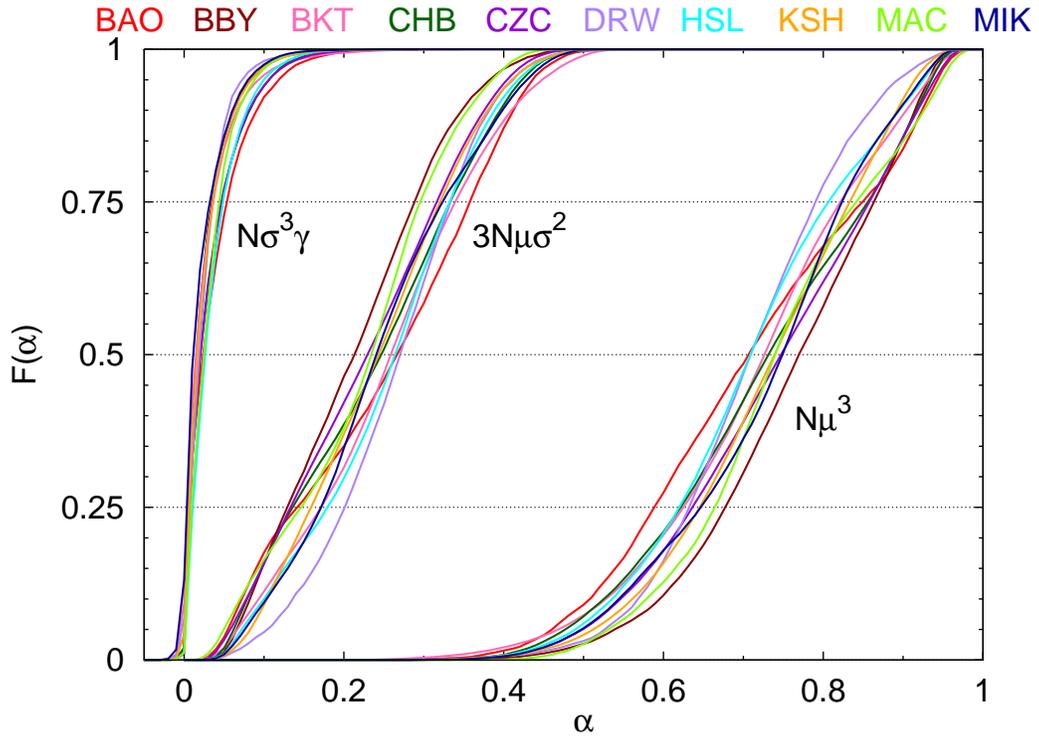}
  \caption{The probability $F(\alpha)$ of contributing a fraction $\alpha$ to the rainfall rate for the three 
factors $N\mu^{3}$, $3N\mu\sigma^{2}$, and $N\sigma^{3}\gamma$ of Eq.(\ref{eq:R_mitica}). Different colors
indicate different data bases.}\label{f5}
\end{figure}

\clearpage

\end{document}